\documentclass[12pt,letterpaper]{article}
\usepackage{osajnl}
\usepackage{overcite,hyperref}

\font\boldsym=cmmib10
\newcommand      \balpha {{\hbox{\boldsym\char'013}}}    %bold \alpha
\def	\ltsim {\leq}       % temporary
\def	\gtsim {\geq}       % temporary

\begin{document}

\title{Discrete-dipole approximation with polarizabilities that account for 
both finite wavelength and target geometry}

\author{Matthew J. Collinge and B. T. Draine}
\address{Princeton University Observatory, Princeton, New Jersey 08544-1001}

\begin{abstract}
The discrete-dipole approximation (DDA) is a powerful method for
calculating absorption and scattering by targets that have sizes
smaller than or comparable to the wavelength of the incident radiation.  
We present a new prescription -- the Surface-Corrected Lattice
Dispersion Relation (SCLDR) -- for assigning the dipole
polarizabilities that takes into account 
both target geometry and finite wavelength.  
We test the SCLDR in DDA calculations using spherical and ellipsoidal 
targets and show that for a fixed number of 
dipoles, the SCLDR prescription results in increased accuracy in the
calculated cross sections for absorption and scattering.
We discuss extension of the SCLDR prescription to irregular targets. 
\end{abstract}

% REPLACE WITH CORRECT OCIS CODES FOR YOUR ARTICLE
                          % NOTE: \ocis{} IS ALIASED TO \pacs{} BUT MUST
                          % FORMAT THE TERMS CORRECTLY FOR EACH JOURNAL
\section{Introduction}

The discrete-dipole approximation (DDA) is a numerical technique for 
calculating scattering and absorption of electromagnetic radiation by targets 
with sizes smaller than or comparable to the incident wavelength. The 
method consists of approximating the target by an array of polarizable 
points (dipoles), assigning polarizabilities at these locations based on the 
physical properties of the target, and solving self-consistently for the 
polarization at each location in the presence of an incident radiation 
field. This procedure can yield arbitrarily accurate results as the number 
of dipoles used to approximate the target is increased. However, computational 
considerations limit the number of dipoles that can be used. Hence, methods 
for increasing the accuracy for a fixed number of dipoles are desirable.

A key factor in determining the level of accuracy that can be reached for a 
given number of dipoles is the prescription 
for assigning dipole polarizabilities. 
In this work, we present a new polarizability prescription that takes into 
account both target geometry and the finite wavelength of incident radiation. 
We test this technique in calculations of absorption and scattering by 
spherical and ellipsoidal targets and show that for a fixed number of dipoles, 
it generally provides 
increased accuracy over previous methods. 
In Section~\ref{pol} we discuss previous 
polarizability prescriptions and develop the new method. In 
Section~\ref{calc} we present calculations testing the new prescription, 
and in Section~\ref{conclusion} 
we discuss our results.

\section{\label{pol}
	 Polarizability Prescriptions}

A fundamental requirement of the DDA is that the inter-dipole separation $d$ 
be small compared to the wavelength of incident radiation, 
$kd\ltsim 1$,
where $k\equiv \omega/c$ is the wavenumber {\it in vacuo}. 
Here 
we will assume the dipoles to be located on a cubic lattice with lattice 
constant $d$, as this facilitates use of fast-Fourier transform (FFT) 
techniques \cite{good90}.

The first implementations of the DDA \cite{purc73} 
used the so-called Clausius-Mossotti 
relation (CMR) to determine the dipole polarizabilities. 
In this procedure, the 
polarizability
$\alpha$ is given as a function of the (complex) refractive 
index $m$ as
\begin{equation}
\alpha_{\rm CMR}={{3d^3}\over{4\pi}}\left({{m^2 -1}\over{m^2 +2}}\right),
\end{equation}
This approach is valid in the infinite wavelength limit of the DDA, 
$kd\rightarrow 0$.

Draine \cite{drai88} showed that for finite wavelengths, the optical theorem 
requires that the polarizabilities 
include a ``radiative-reaction'' correction of the form
\begin{equation}
\alpha={{\alpha^{(nr)}}\over{1-(2/3)i(\alpha^{(nr)}/d^3)(k d)^3}},
\end{equation}
where $\alpha^{(nr)}$ is the ``non-radiative'' polarizability, 
that is, before any radiative-reaction correction is applied. Draine 
\cite{drai88} used 
$\alpha_{\rm CMR}$
as the non-radiative polarizability.

Based on analysis of an integral formulation of the scattering problem, 
Goedecke \& O'Brien \cite{goed88} and Hage \& Greenberg \cite{hage90} 
suggested further corrections 
to the CMR polarizability of order $(k d)^2$. Draine \& Goodman \cite{drai94} 
studied 
electromagnetic wave propagation on an infinite lattice; they required that 
the lattice reproduce the dispersion relation of a continuum medium. In this 
``Lattice Dispersion Relation'' (LDR) approach, the radiative-reaction 
correction emerges naturally, and the polarizability is given 
[to order $(kd)^3$] by
\begin{equation}
\label{eq:LDR}
\alpha_{\rm LDR}={
{\alpha^{(0)}}
\over{1+(
\alpha^{(0)}
/d^3)[(b_1+m^2 b_2+m^2 b_3 S)(k d)^2-(2/3)i(k d)^3]}},
\end{equation}
where 
$\alpha^{(0)}=\alpha_{\rm CMR}$ 
is the polarizability in the limit $kd\rightarrow 0$,
$b_1=-1.8915316$, $b_2=0.1648469$ and $b_3=-1.7700004$, 
and $S$ is a function of the propagation direction and polarization of the 
incident wave. $S$ is given as 
\begin{equation}
\label{eq:S}
S=\sum_j (a_j e_j)^2,
\end{equation}
where $\bf a$ and $\bf e$ are the unit propagation and polarization vectors, 
respectively. 
Note that 
eq.\ (\ref{eq:S}) gives
$S=0$ for waves propagating along any of the lattice axes.
This method correctly accounts to $O[(k d)^3]$ for the 
finite wavelength of incident radiation, and by construction, it accurately 
reproduces wave propagation in an infinite medium. Its primary limitation is that 
the accuracy in computing absorption cross-sections of finite targets 
(for a given number of dipoles) degrades rapidly as 
the imaginary part of the refractive index $m$ becomes large (e.g., for 
${\mathrm {Im}}(m)
\gtsim 2$). 

\subsection{Geometric Corrections: the Static Case}

Recently Rahmani, Chaumet \& Bryant \cite{rahm02} (RCB) 
proposed a new method for 
assigning the polarizabilities that takes into account the effects of target 
geometry on the local electric field at each dipole site. 
Consider a continuum target in a static, uniform applied field ${\bf E}^0$.
At each location $i$ in the target, the macroscopic electric
field ${\bf E}_i^m$ is linearly related to ${\bf E}^0$:
\begin{equation}
\label{eq:Em_and_E0}
{\bf E}_i^m = {\bf C}_i^{-1} {\bf E}^0
\end{equation}
where ${\bf C}_i^{-1}$ is a $3\times 3$ tensor 
that will depend on location $i$, 
the global geometry of the
target, and its (possibly nonuniform) composition.
If we now represent the target by a dipole array, and require that the
electric dipole moment ${\bf P}_i$ of dipole $i$ 
be equal to $d^3$ times the
macroscopic polarization density at location $i$, we obtain
\begin{equation}
\label{eq:P_and_E0}
{\bf P}_i = d^3 \left(\frac{\epsilon_i-1}{4\pi}\right){\bf E}_i^m
= d^3 \left(\frac{\epsilon_i-1}{4\pi}\right){\bf C}_i^{-1}{\bf E}^0
\end{equation}
If $\balpha_i$ is the polarizability tensor of dipole $i$,
then
\begin{equation}
\label{eq:P_and_alpha}
{\bf P}_i = 
\balpha_i\left[{\bf E}^0 - \sum_{j\neq i}{\bf A}_{ij}{\bf P}_j\right]
\end{equation}
where $-{\bf A}_{ij}{\bf P}_j$ is the contribution to the electric field
at location $i$ due to dipole ${\bf P}_j$ at location $j$
(this defines the 3$\times$3 tensors ${\bf A}_{ij}$).
Substituting (\ref{eq:P_and_E0}) into (\ref{eq:P_and_alpha})
we obtain
\begin{equation}
\balpha_i = d^3\left(\frac{\epsilon_i-1}{4\pi}\right){\bf \Lambda}_i^{-1}
\end{equation}
where the 3$\times$3 tensors
\begin{equation}{\bf \Lambda}_i \equiv {\bf C}_i-
\sum_{j\neq i}{\bf A}_{ij}\left(\frac{\epsilon_j-1}{4\pi}\right)d^3
{\bf C}_j^{-1}{\bf C}_i
\end{equation}
can be evaluated (and easily inverted)
if the ${\bf C}_i$ are known.

The RCB approach requires that
the tensors ${\bf C}$ first be obtained. 
For certain simple geometries, the ${\bf C}_i$ can be obtained analytically.
For example, for homogeneous ellipsoids,
infinite slabs, or infinite cylinders, the tensors ${\bf C}_i$ can
be expressed in the form
\begin{equation}
{\bf C}_i = 1 + \left(\frac{\epsilon-1}{4\pi}\right){\bf L}
\end{equation}
where ${\bf L}$ is a ``depolarization tensor''.
For example, $L=1/3$ for a homogeneous sphere.

In the present work, we combine the LDR and RCB approaches 
in order to obtain a polarizability prescription that accounts both 
for finite wavelength and for local field corrections arising from target 
geometry. 
We adopt $\balpha_{\rm RCB}$ as the 
polarizability $\alpha^{(0)}$ in the limit $kd\rightarrow0$,
and apply 
corrections up to $O[(k d)^3]$ based on the LDR. A further analysis of 
the electromagnetic dispersion relation of a non-cubic lattice \cite{report} called 
into question the value of the constant $b_3$ in eq.~(\ref{eq:LDR}) used by 
Draine \& Goodman \cite{drai94}, 
and found it instead to be undetermined by available constraints. Thus 
we include 
an adjustable
factor $f$ whose value is chosen to optimize 
the behavior of the new method as discussed in the next section. 
The ``Surface-Corrected Lattice Dispersion Relation'' (SCLDR) polarizability 
is given by 
\begin{equation}
{\bf \balpha}_{\rm SCLDR}={\bf \balpha}_{\rm RCB}\{1+({\bf \balpha}_{\rm RCB}/d^3)[(b_1+{\bf m}^2 b_2+{\bf m}^2 b_3 f S)(k d)^2-(2/3)i(k d)^3]\}^{-1},
\end{equation}
where
\begin{equation}
f=\exp[-0.5{\rm Im}(m)^2].
\label{eq:fittingf}
\end{equation}
In the next section, 
we test this new prescription in calculations of absorption 
and scattering by spherical and ellipsoidal targets.

\section{\label{calc}
         Sphere and Ellipsoid Calculations}

For a continuum target of volume $V$, the effective radius
$a_{\rm eff}\equiv (3V/4\pi)^{1/3}$, the radius of a sphere of equal volume.
The target is approximated by an array of $N$ dipoles located on a cubic
lattice, with the dipole locations selected by some criterion designed
to approximate the shape of the original target.
The inter-dipole spacing is then set to $d=(V/N)^{1/3}$.

For a given orientation of the dipole array relative to the incident wave,
we calculate the cross sections $C_{\rm sca}$ and $C_{\rm abs}$ for
scattering and absorption, and the dimensionless efficiency factors
$Q_{\rm sca}\equiv C_{\rm sca}/\pi a_{\rm eff}^2$,
$Q_{\rm abs}\equiv C_{\rm abs}/\pi a_{\rm eff}^2$.

To test the performance of the SCLDR polarizability prescription 
against previous results, we 
performed a series of calculations using the 
DDA code DDSCAT \cite{drai00}, 
modified to permit use of the SCLDR polarizabilities. 
We computed 
$Q_{\rm sca}$ and $Q_{\rm abs}$
for spherical 
targets with a range of refractive indexes and for a range of scattering 
parameters $x=2\pi a_{\rm eff}/\lambda=ka_{\rm eff}$, 
using three different approaches 
for assigning the dipole polarizabilities: 
LDR, RCB and SCLDR. 
Spherical targets 
were employed because the exact optical properties can be 
readily calculated using Mie theory. 
We also performed a similar but more 
limited set of calculations for 
ellipsoidal targets.

We tested the LDR, RCB and SCLDR prescriptions for a number of different 
refractive indexes in the region of the complex plane with ${\rm Re}(m)\leq 5$ 
and ${\rm Im}(m)\leq 4$. 
We determined that for refractive indexes with these ranges 
of real and imaginary parts, it was desirable for the SCLDR correction 
factor $f$ to tend toward unity 
for ${\rm Im}(m)<1$ and to tend toward zero 
for ${\rm Im}(m)>2$. We chose the functional form of eq.~(\ref{eq:fittingf}) 
in order to reproduce this asymptotic behavior. 

\begin{figure}[ht]
   \centerline{
   \scalebox{0.7}{
   \includegraphics{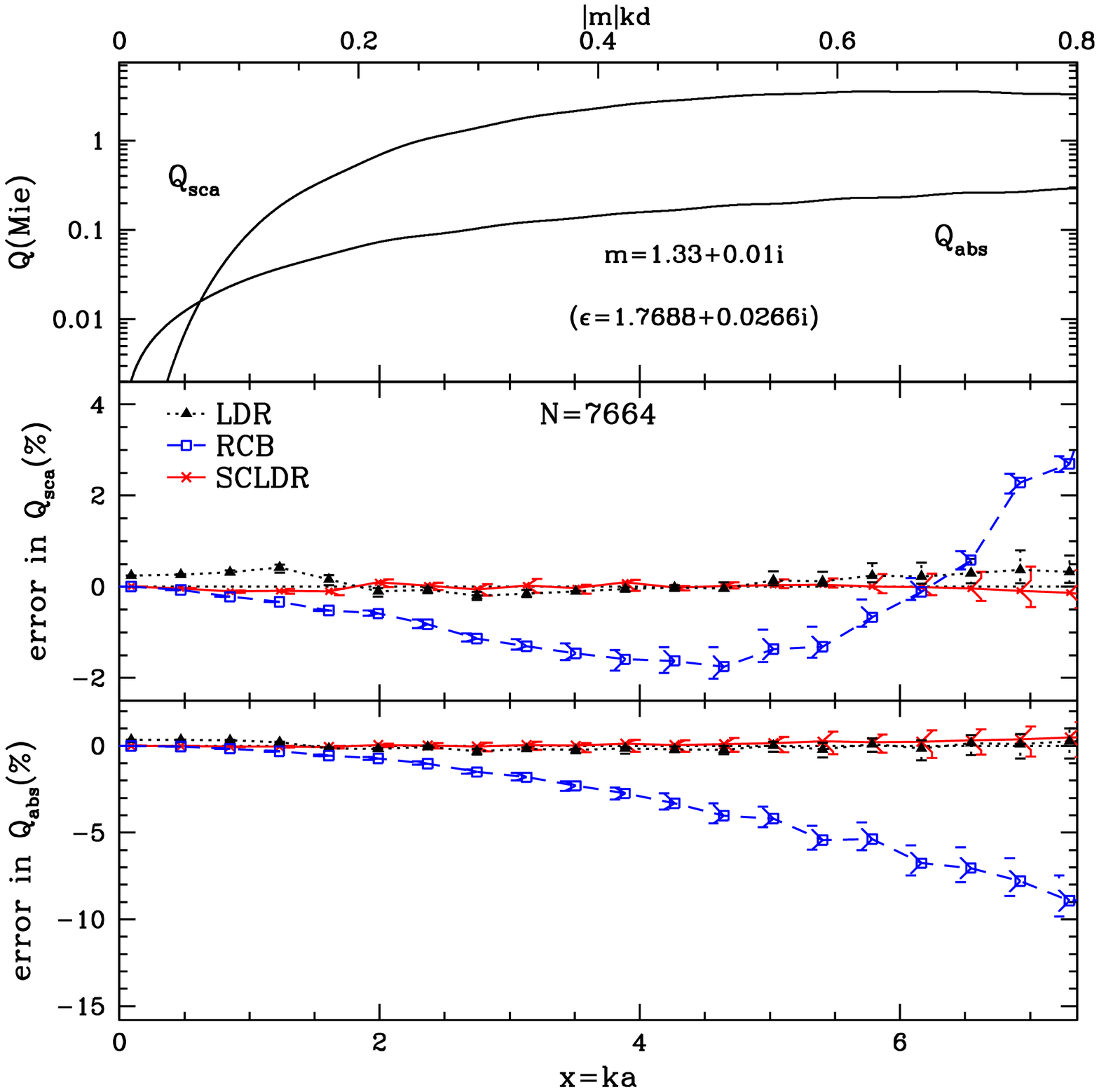}}}
\caption{\label{sphere133}
   Comparison of scattering and absorption 
efficiency factors 
$Q_{\rm sca}$ and $Q_{\rm abs}$ computed for a pseudo-sphere 
of $N=7664$ dipoles and 
refractive index $m=1.33+0.01i$, averaged over 12 orientations, and 
using three different polarizability 
prescriptions: Lattice Dispersion Relation (LDR); Rahmani et~al. \cite{rahm02}
(RCB); 
and Surface-Corrected Lattice Dispersion Relation (SCLDR). 
The horizontal axis shows (top) $|m|kd$ (the phase shift in radians within one 
lattice spacing) and (bottom) the scattering parameter $x=ka$. 
Error bars indicate the 
ranges of $Q$ values obtained for the individual orientations. 
The top panel shows 
the results of Mie theory calculations; the lower panels show the fractional 
error in $Q_{\rm sca}$ and $Q_{\rm abs}$, respectively. 
The SCLDR and LDR prescriptions 
are clearly preferred over RCB for this case.
} 
\end{figure}

\begin{figure}[ht]
   \centerline{
   \scalebox{0.7}{
   \includegraphics{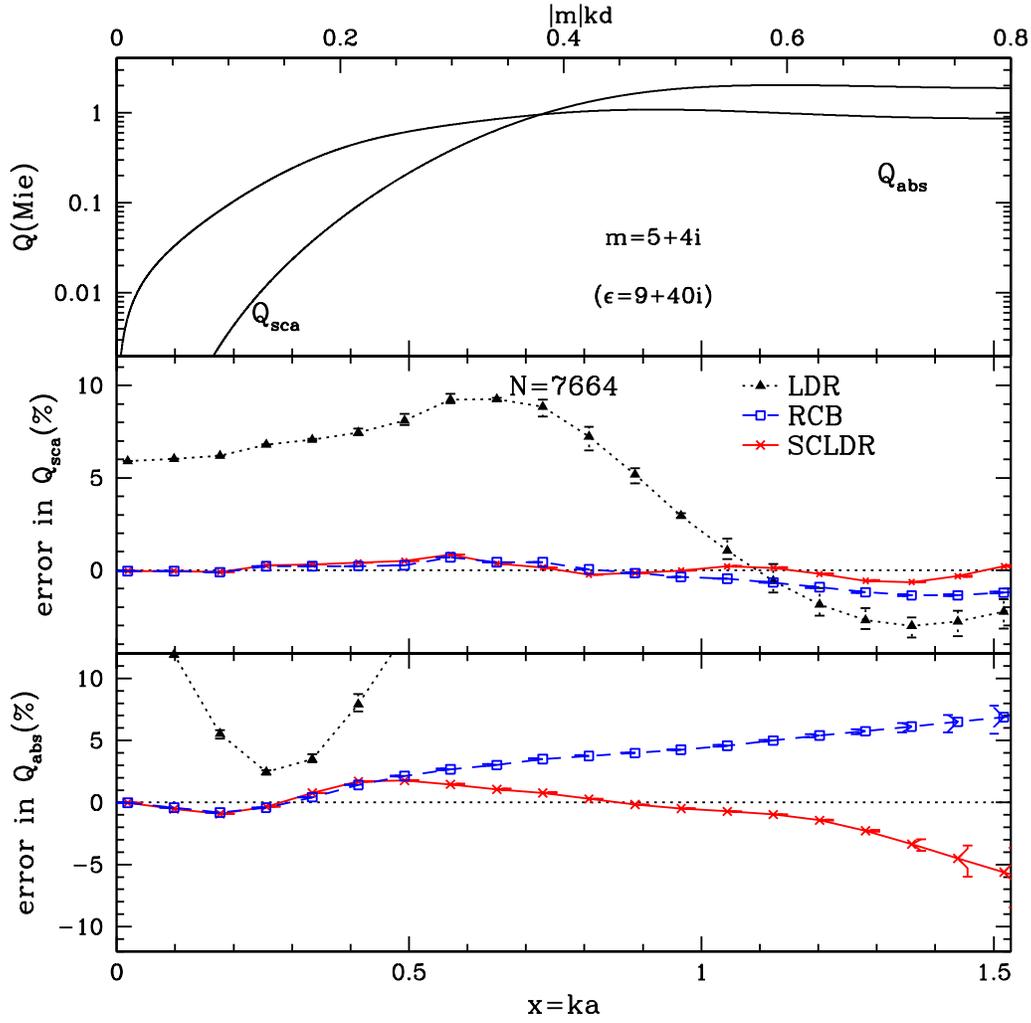}}}
\caption{\label{sphere54}
Same as Fig.~\ref{sphere133}, but for refractive 
index $m=5+4i$. 
The SCLDR and RCB prescriptions are clearly preferred over the 
LDR for this case, with SCLDR being somewhat superior to RCB.
} 
\end{figure}

Figures \ref{sphere133} and \ref{sphere54} show the results of calculations 
for spheres with refractive indices $m=1.33+0.01i$ and $m=5+4i$, each
approximated by an array of $N=7664$ dipoles.
Because the dipole array is not rotationally symmetric, $Q_{\rm sca}$ and
$Q_{\rm abs}$ calculated with the DDA depend in general on the target orientation;
we perform calculations for 12 orientations, and we show the average and
range of the results.
We calculate the fractional errors in 
$Q_{\rm sca}$ and $Q_{\rm abs}$ by comparison with 
exact results obtained using Mie theory:
\begin{equation}
{\rm frac.err}\equiv \frac{Q({\rm DDA})-Q({\rm Mie})}{Q({\rm Mie})} ~~~.
\end{equation}

In previous work\cite{drfl94} it was recommended that the DDA be used only when
$|m|kd \ltsim 1$, or a more stringent condition $|m|kd < 0.5$ if the
DDA is to be used to calculate the differential scattering cross section.
In the present work we find that when the SCLDR polarizabilities are used,
the fractional errors in $Q_{\rm sca}$ and $Q_{\rm abs}$ 
are relatively insensitive to $x$ provided
$|m|kd \leq 0.8$, which we adopt as an operational validity criterion.
Figures \ref{sphere133} and \ref{sphere54} show results for
values of $x$ satisfying 
$|m|kd \leq 0.8$.

From Figure~\ref{sphere133}, it is clear that the LDR and SCLDR prescriptions 
provide approximately equal levels of accuracy in the $|m|\approx 1$ regime, 
while the RCB prescription does not perform as well. Figure~\ref{sphere54} 
shows that at the other extreme of ${\rm Re}(m)\gg 1$ and ${\rm Im}(m)\gg 1$, 
the LDR approach results in large errors, especially in the calculated 
absorption cross sections, while the RCB and SCLDR prescriptions perform 
approximately equally well.

\begin{figure}[ht]
   \centerline{
   \scalebox{0.7}{
   \includegraphics{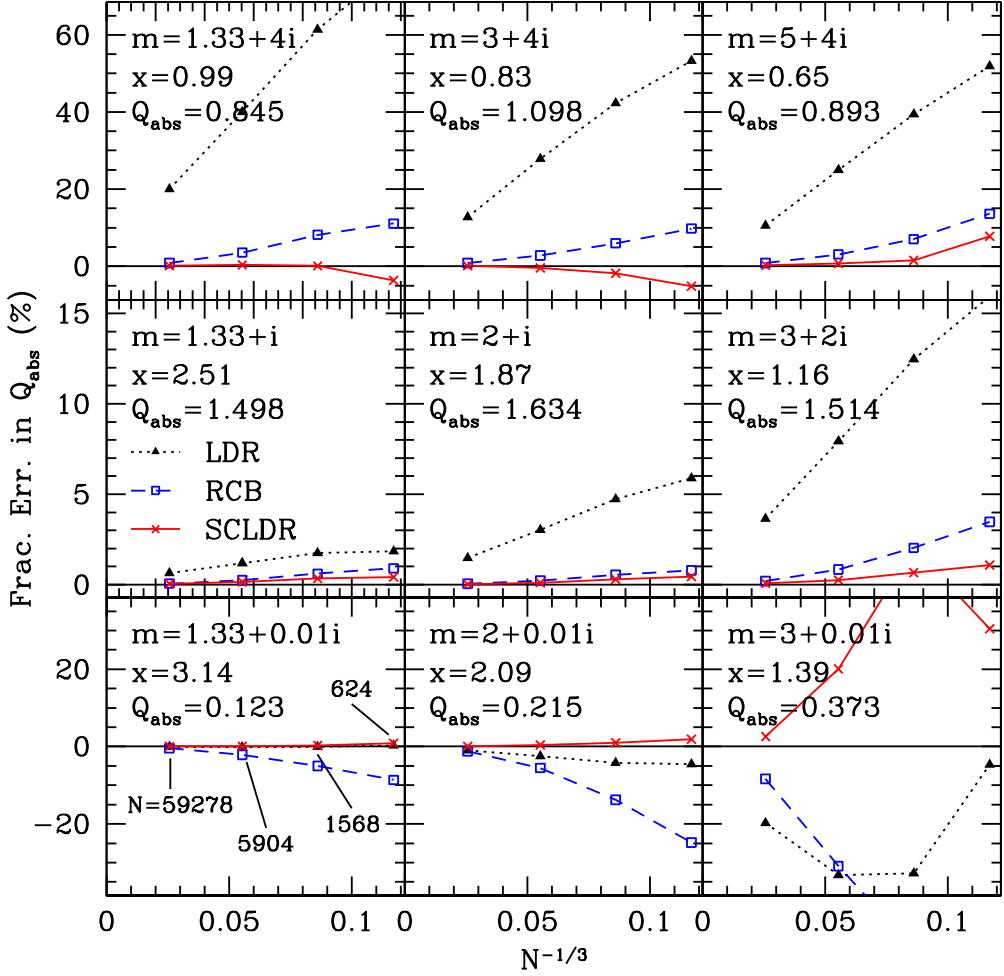}}}
\caption{\label{converge_sphere_abs}
Fractional error in $Q_{\rm abs}$ 
averaged over 12 orientations for spheres with 
different refractive indexes, as a function of $N^{-1/3}$, where $N$ is the 
number of dipoles, in the range 624--59278. Calculations are shown for the 
LDR, RCB, 
and SCLDR polarizability prescriptions; the symbolic 
scheme is the same as in Fig.~\ref{sphere133}. Refractive indexes $m$,
scattering parameters $x=ka$, and exact values of $Q_{\rm abs}$ computed from Mie 
theory are shown in the left portion of each panel. 
The scattering parameters are chosen so that $|m|kd\approx 0.8$ 
(the approximate 
limit of applicability of the DDA) for the smallest number ($N=624$) 
of dipoles. The convergence with 
increasing $N$ is quite smooth in all regions of the complex $m$-plane, with 
the exception of $m=3+0.01i$. In almost every case shown, 
fractional errors $<2$\% (and often significantly lower) can be achieved for 
$N\approx 6000$ dipoles. We find that for calculating $Q_{\rm abs}$, 
the SCLDR is comparable or superior in 
accuracy to the LDR and RCB prescriptions throughout the region of $m$-space 
shown.}
\end{figure}

\begin{figure}[ht]
      \centerline{
      \scalebox{0.7}{
      \includegraphics{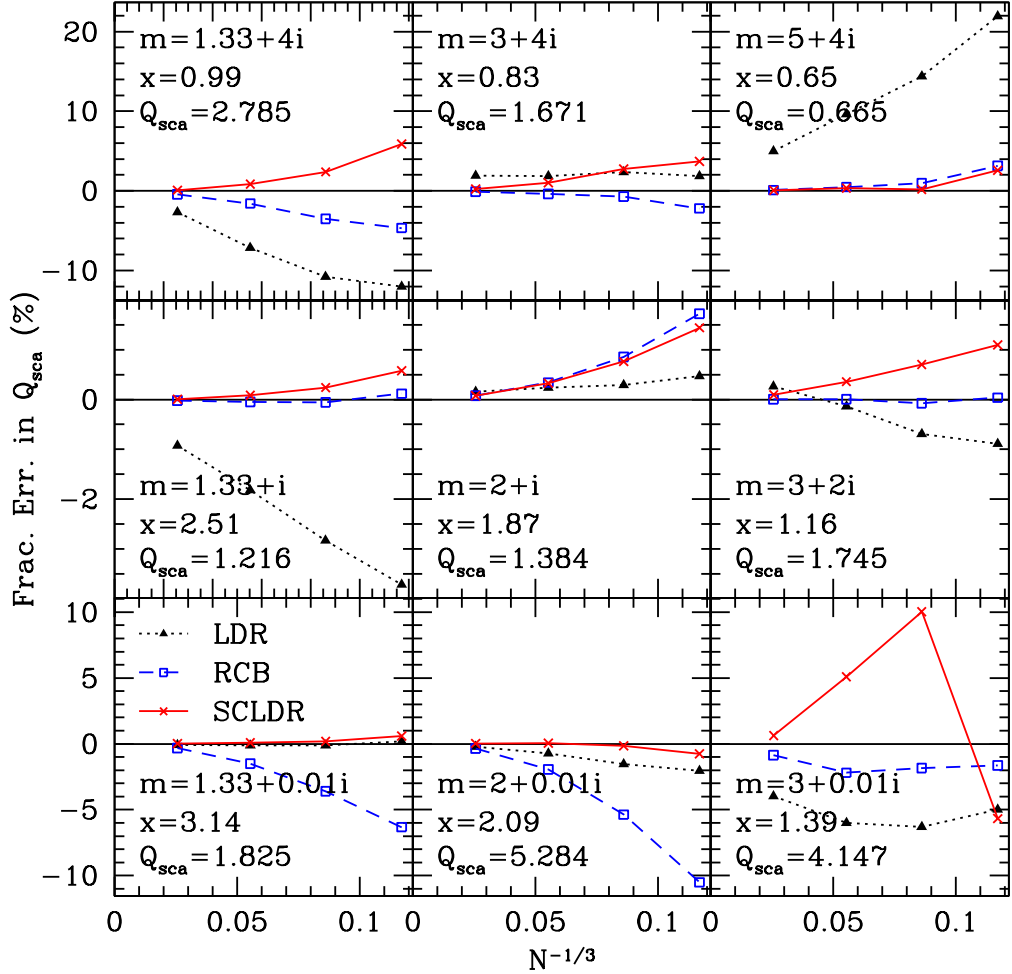}}}
\caption{\label{converge_sphere_sca}
Same as Fig.~\ref{converge_sphere_abs}, 
except that fractional errors in $Q_{\rm sca}$ are plotted. Again, the 
SCLDR prescription is comparable or superior to the LDR prescription 
for all values of $m$ shown. While the 
SCLDR prescription is still comparable or superior to the RCB prescription for 
$m$ values with small imaginary parts, the RCB prescription provides better 
accuracy in calculating $Q_{\rm sca}$ for $m$ values with large imaginary parts. 
}
\end{figure}

In Figures~\ref{converge_sphere_abs} and~\ref{converge_sphere_sca}, 
we show the convergence behavior of the 
different polarizability prescriptions as the 
number of dipoles $N$ is increased for spherical targets with selected
refractive indices;
the refractive indices have been chosen to
sample the region of the complex plane discussed in the previous 
paragraphs. 
The SCLDR method performs comparably to or 
better than the RCB and LDR prescriptions throughout this region of the 
complex refractive index plane. This illustrates the advantage of the SCLDR 
approach over these previous techniques: it performs well not just for a 
small range of refractive indexes, but for the entire range we have sampled.

\begin{figure}[ht]
   \centerline{
   \scalebox{0.7}{
    \includegraphics{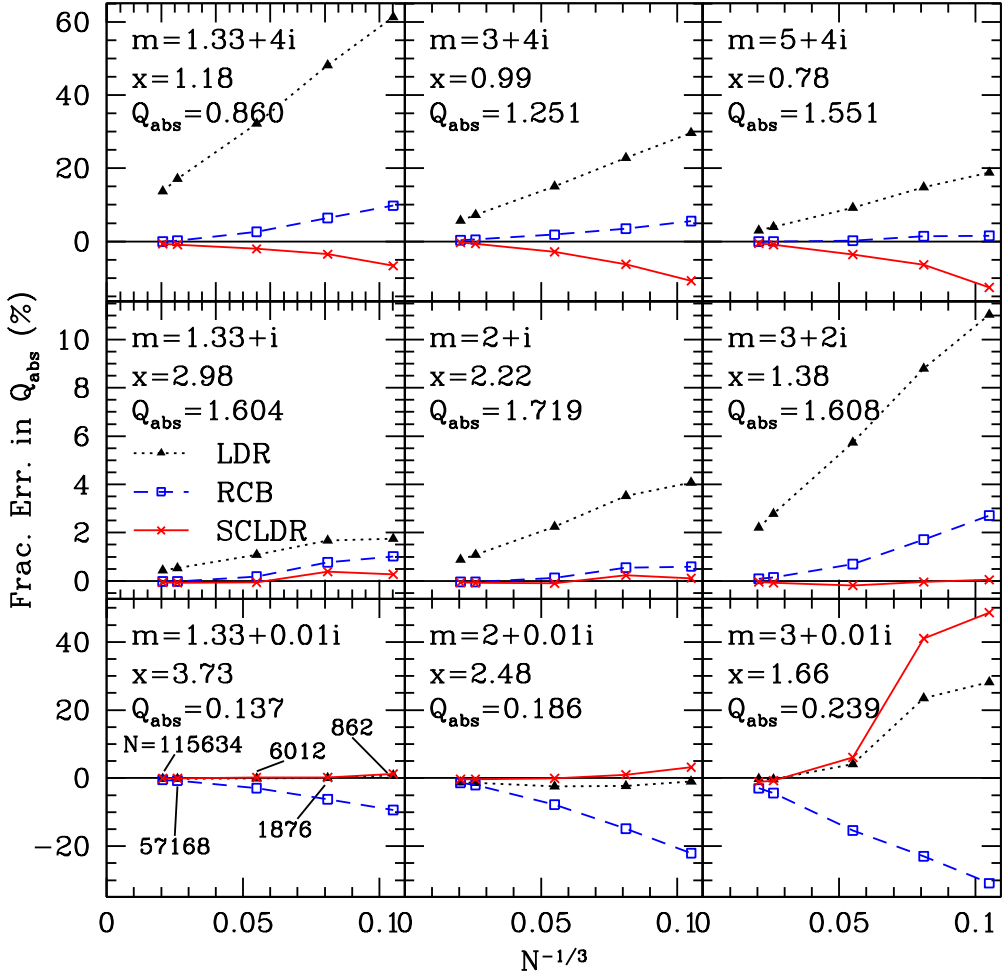}}}
\caption{\label{converge_ellipse_abs}
Same as Fig.~\ref{converge_sphere_abs}, but for ellipsoids with
approximately 1:2:3 axial ratios.  
Fractional errors have been
estimated based on comparison with an extrapolation of the convergence
behavior of the three polarizability prescriptions, as described in
Section~\ref{calc}.
Again the SCLDR prescription appears comparable or superior to the LDR
and RCB prescriptions for calculating $Q_{\rm abs}$ throughout the region
of the complex $m$-plane sampled, though the RCB prescription is
slightly preferred for the cases of $m=3+4i$ and $m=5+4i$.
}
\end{figure}

\begin{figure}[ht]
    \centerline{
    \scalebox{0.7}{
    \includegraphics{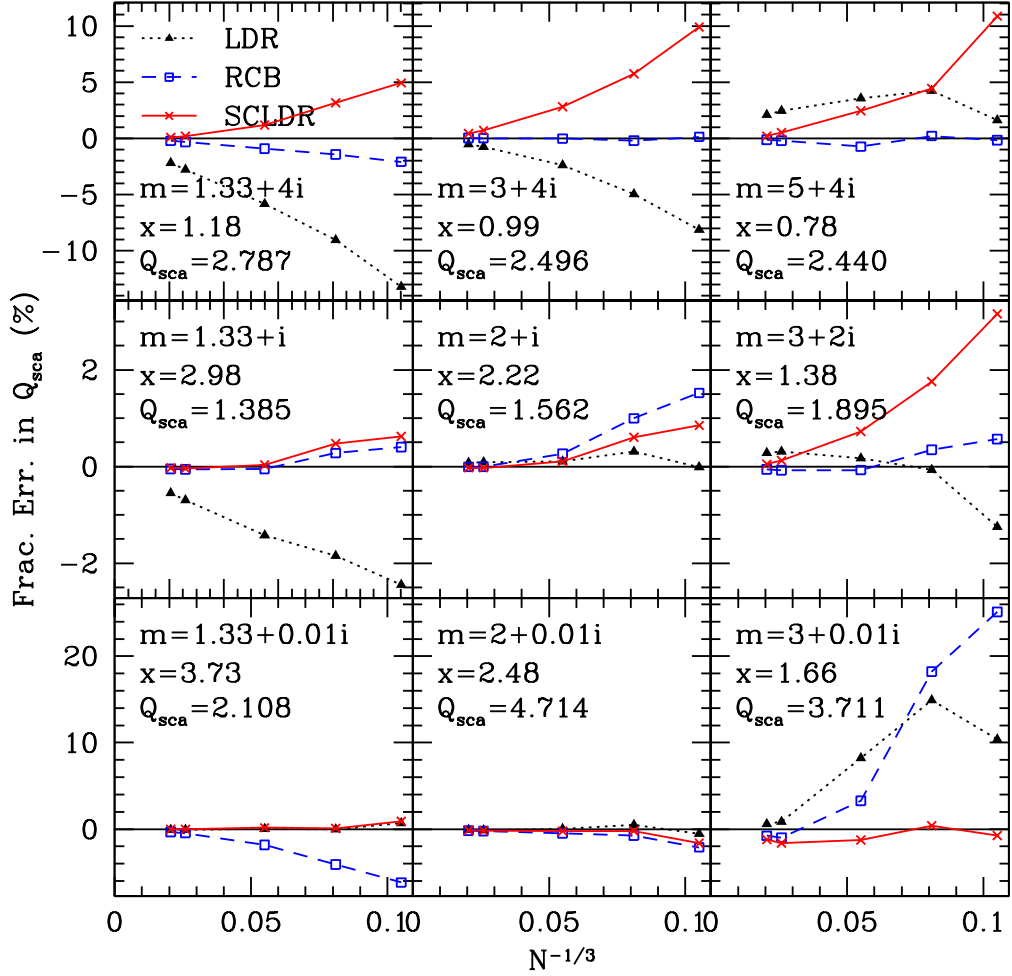}}}
\caption{\label{converge_ellipse_sca}
Same as Fig.\ \ref{converge_ellipse_abs}, except that fractional errors
in $Q_{\rm sca}$ are plotted.
As in Fig.\ \ref{converge_sphere_abs}, SCLDR is comparable to or superior
to LDR for all values of $m$, and to RCB for values of $m$ with small
imaginary parts, while RCB is somewhat superior to SCLDR for values
of $m$ with large imaginary parts.}
\end{figure}

Figures~\ref{converge_ellipse_abs} and~\ref{converge_ellipse_sca} extend 
the result shown in Figures~\ref{converge_sphere_abs} and~\ref{converge_sphere_sca} 
to targets of a more general shape, specifically 
ellipsoids with approximately 1:2:3 axial ratios. For these targets, 
we have estimated the true values of 
$Q_{\rm sca}$ and $Q_{\rm abs}$ by 
assuming these to be linear functions of $N^{-1/3}$, extrapolating to 
$N^{-1/3}\rightarrow 0$ for each polarizability prescription, 
and taking the average of the 
results from the different prescriptions. The close similarity of 
the results of these calculations to those shown in 
Figures~\ref{converge_sphere_abs} and~\ref{converge_sphere_sca} 
demonstrates that the SCLDR prescription provides 
the same benefits in calculations for ellipsoidal targets as for spheres, 
although we note that for ellipsoids with 
values of $m$ with large imaginary parts 
[typically ${\rm Im}(m)>1$], 
the RCB prescription can provide improved
accuracy in calculations of $Q_{\rm sca}$.

\begin{figure}[ht]\centerline{
       \rotatebox{270}{
       \scalebox{0.6}{
       \includegraphics{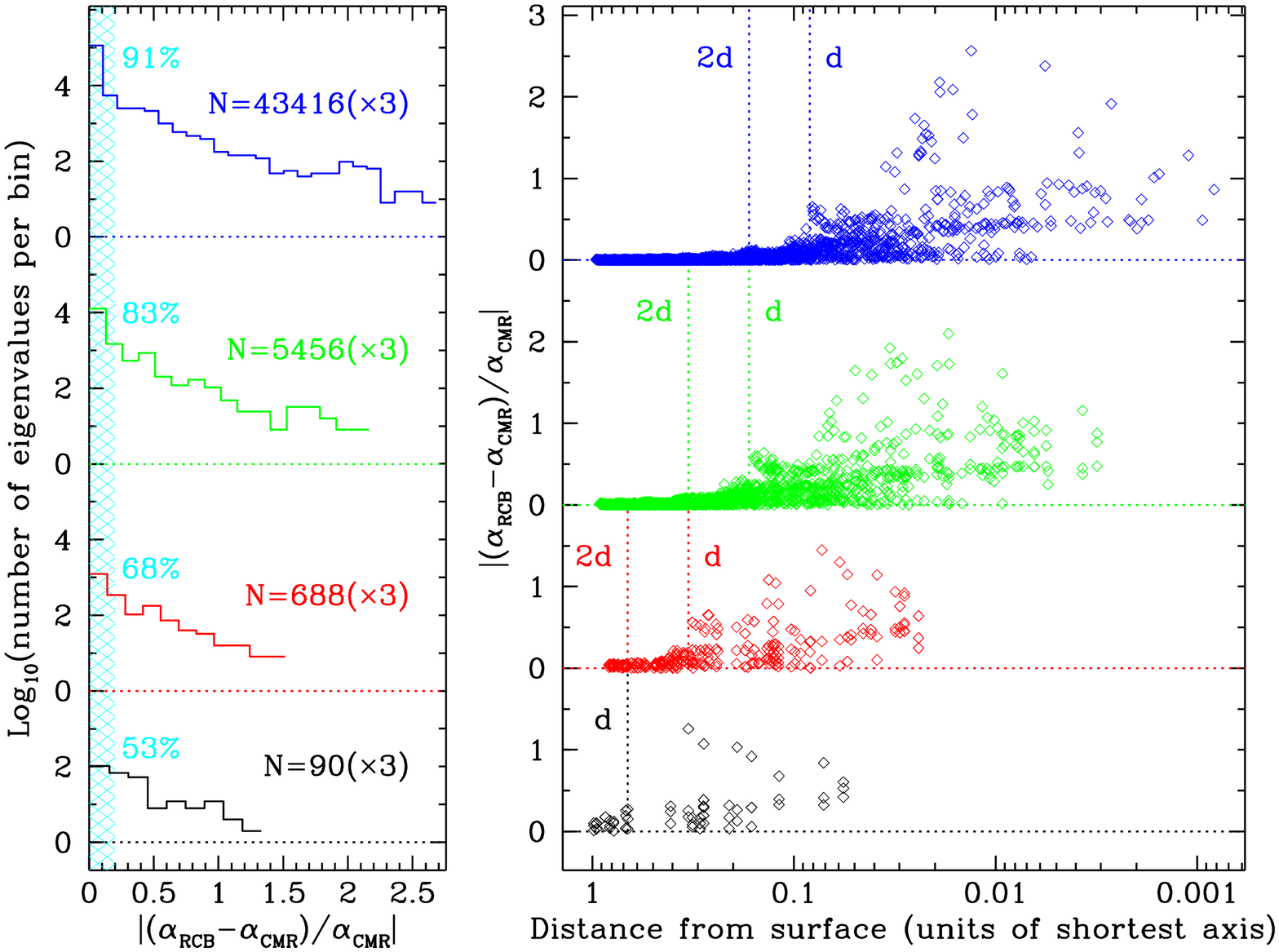}}}}
\caption{\label{fig:cmtest}
        Comparison of RCB and CMR polarizabilities. 
	The left panel shows the
	distribution of polarizability eigenvalues
	for discrete dipole approximations to
	a 1:2:3 ellipsoid with $m=5+4i$ 
	using $N=90$, 688, 5456, and 43416 
	dipoles.
	The shaded region corresponds
        to a fractional difference of 20\% or less; the fraction of the
	eigenvalues within this region varies from 53\% for $N=90$
	($3d\times6d\times9d$ axes) to
	91\% for $N=43416$ 
	($24d\times48d\times72d$ axes).
	The right panel shows the
	fractional difference between RCB and CMR polarizabilities versus the
	distance (in units of the shortest axis) 
	from the ideal ellipsoidal surface used to define the target 
	(all dipole locations are
	interior to this surface).
	As expected, 
	the RCB polarizability reduces to the CMR polarizability for dipoles
	lying more than $\sim$2$d$ from the surface.
}
\end{figure}

For an isotropic material with refractive index $m$, 
the Clausius-Mossotti polarizability 
$\alpha_{\rm CMR}$ has triply-degenerate eigenvalues $\alpha_{\rm CMR}=
(m^2-1)d^3/4\pi$.
For the case of a 1:2:3 ellipsoid with refractive index
$m=5+4i$, we have calculated the eigenvalues $\alpha$
of $\balpha_{\rm SCLDR}$ for $|m|kd\rightarrow 0$ 
(for which case $\balpha_{\rm SCLDR}\rightarrow \balpha_{\rm RCB}$)
at each occupied lattice site.
Figure \ref{fig:cmtest} (left panel) shows the distribution of the
fractional difference of the eigenvalues $\alpha$ from $\alpha_{\rm CMR}$.
The deviations tend to be appreciable 
(fractional difference exceeding $\sim$20\%)
only near the surface.
The left panel shows that the deviations exceed 20\% for 47\% of the lattice
sites for $N=90$, but only  9\% of the lattice sites when $N=43416$.
For this example
the fraction of the eigenvalues deviating by
$>$20\% is $\sim 0.30 (N/1000)^{-1/3}$ for $N \gtsim 500$, 
approximately equal 
to the fraction of the dipoles 
located within
a surface layer of thickness $\sim0.6d$.

The right panel in Figure \ref{fig:cmtest} shows the eigenvalue
deviations as a function of distance from the surface of the ellipsoid: the
eigenvalues deviating from $\alpha_{\rm CMR}$
by more than $\sim20\%$ are, as expected, exclusively
associated with dipoles located within a distance $d$ of the surface.

\section{\label{conclusion}
         Conclusion}

We introduce a new DDA polarizability prescription -- the Surface-Corrected 
Lattice Dispersion Relation (SCLDR). This technique builds on previous work, 
principally by Draine \& Goodman \cite{drai94} 
and Rahmani, Chaumet \& Bryant \cite{rahm02}, to 
account properly for both finite wavelength and target geometry. We have tested 
the new polarizability prescription in calculations of absorption and 
scattering by spherical and ellipsoidal targets. These tests show that 
the SCLDR performs generally better than previous prescriptions which took 
account either of finite wavelength or of target geometry but not both. The 
SCLDR technique is most easily applicable to target shapes for which there 
exists an analytical solution to the electrostatic applied field problem, but 
it can be applied to any dielectric target (homogeneous or inhomogeneous, 
isotropic or anisotropic) provided that the electrostatic problem can at 
least be solved numerically to obtain the tensors ${\bf C}_i$ 
(see eq.~\ref{eq:Em_and_E0}). In such cases, it generally provides a 
significant increase in accuracy over previous methods, especially for 
highly absorptive materials.

%After the manuscript is proofread, the {\tt .tex} file and figures
%should be tarred and gzipped.  Follow the instructions on the OSA
%Publications web site for submitting through the e-subs system
%(\url{http://www.osa/org/pubs}). Authors should feel free to
%contact OSA staff for assistance (see appropriate journal page on
%the web site for contact information). \TeX-specific questions or
%suggestions may be directed to Scott Dineen ({\tt sdinee@osa.org})
%or Chris Mayfield ({\tt cmayfi@osa.org}).

\section*{Acknowledgments}
This research was supported in part by NSF grant AST-9988126. M.J.C. also
acknowledges support from a NDSEG Fellowship. The authors wish to thank
Robert Lupton for making available the SM software package.

\newpage

\begin{thebibliography}{}
%%Do not include separate BibTeX files; if BibTeX is used, paste the output here.

\bibitem[1]{good90}
J. J. Goodman, B. T. Draine, and P. J. Flatau, ``Application of fast-Fourier 
transform techniques to the discrete dipole approximation,'' Opt. Lett. 16, 
1198--1200 (1990).
\bibitem[2]{purc73}
E. M. Purcell and C. R. Pennypacker, ``Scattering and absorption of light 
by nonspherical dielectric grains,'' Astrophys. J. 186, 705--714 (1973).
\bibitem[3]{drai88}
B. T. Draine, ``The discrete-dipole approximation and its application to 
interstellar graphite grains,'' Astrophys. J. 333, 848--872 (1988).
\bibitem[4]{goed88}
G. H. Goedecke and S. G. O'Brien, ``Scattering by irregular inhomogeneous 
particles via the digitized Green's function algorithm,'' Appl. Opt. 27, 
2431--2438 (1988).
\bibitem[5]{hage90}
J. I. Hage and J. M. Greenberg, ``A model for the optical properties of porous 
grains,'' Astrophys. J. 361, 251--259 (1990).
\bibitem[6]{drai94}
B. T. Draine and J. Goodman, ``Beyond Clausius-Mossotti: wave propagation on 
a polarizable point lattice and the discrete dipole approximation,'' Astrophys. 
J. 405, 685--697 (1993).
\bibitem[7]{rahm02}
A. Rahmani, P. C. Chaumet, and G. W. Bryant, ``Coupled dipole method with an 
exact long-wavelength limit and improved accuracy at finite frequencies,'' Opt. 
Lett. 27, 2118--2120 (2002).
\bibitem[8]{report}
D. Gutkowicz-Krusin and B. T. Draine, in preparation
\bibitem[9]{drai00}
B. T. Draine and P. J. Flatau, ``User Guide for the Discrete Dipole Approximation 
Code DDSCAT (Version 5a10),'' http://xxx.arXiv.org/abs/astro-ph/0008151v3, 1--42 
(2000).
\bibitem[10]{drfl94}
B. T. Draine and P. J. Flatau,
``The discrete dipole approximation for scattering calculations'',
J. Opt. Soc. Am. A, 11, 1491--1499 (1994).

\end{thebibliography}
\end{document}